# Acoustic Metagrating Circulators: Nonreciprocal, Robust, and Tunable Manipulation with Unitary Efficiency

Lijuan Fan, and Jun Mei*

*School of Physics, South China University of Technology, Guangzhou 516040, China.*

**Abstract**

Nonreciprocal signal operation is highly desired for various acoustic applications, where protection from unwanted backscattering can be realized so that transmitting and receiving signals are processed in a full-duplex mode. Here we present the realization of a class of nonreciprocal circulators based on simply structured acoustic metagratings, which consist only of a few solid cylinders and a steady fluid flow with low velocity. These innovative metagratings are intelligently designed via a diffraction analysis of the linearized potential flow equation and a genetic-algorithm-based optimization process. Unitary reflection efficiency between desired ports of the circulators are demonstrated through full-wave numerical simulations, confirming nonreciprocal and robust circulation of the acoustic signal over a broad range of flow velocity magnitude and profile. Our design provides a feasible degree of tunability, including switching from reciprocal to nonreciprocal operation and reversing the handedness of the circulator, presenting a convenient but efficient approach for the realization of nonreciprocal acoustic devices from wavelength-thick metagratings. It may find applications in various scenarios including underwater communication, energy harvesting, and acoustic sensing.

———————
*phjunmei@scut.edu.cn

# I. INTRODUCTION

Reciprocity is a fundamental property in many fields of physics such as acoustic, mechanics, and electromagnetics, which means that if a signal can travel from point A to point B, they can also travel backward from B to A in an identically way [1,2]. In fact, reciprocity is dedicated by the fact that the underlying wave equations (such as the acoustic, electrodynamic or Maxwell's equations) are symmetric with respect to the time variable $t$. Although a lot of technologies and devices rely on the validity of reciprocity, in many scenarios we intendedly break reciprocity to transmit data in a unidirectional, robust, and no-jamming way so that transmitting and receiving signals can be processed in a full-duplex mode. In a multi-port system, nonreciprocity provides protection from unwanted backscattering by the loads, which could protect sensitive equipment and allows components to be added in a modular fashion. Therefore, breaking reciprocity has been an interesting topic of acoustic research in recent years. Generally speaking, nonreciprocal propagation of acoustic waves can be achieved by applying a circulating fluid motion [3], through duct flow inside waveguides [4], via interferences between clockwise/counterclockwise modes of a ring resonator [5], through magnetoelastic coupling [6,7], via dynamic modulation [8-10], or utilizing nonlinearities [11-14].

On the other hand, metagrating, a wavelength-thick artificial structure based on the diffraction theory, is proposed and studied very recently, and has received considerable attention [15–18]. Through a proper design, a discrete set of diffraction orders of the metagrating can be engineered in a controlled way so that the wave energy is directed to desired directions with nearly unitary efficiency, breaking the limitation of the generalized Snell's law [19,20]. One distinguishing characteristic of metagrating, as compared with metasurface, is that people do not need to discrete a fast-varying impedance or phase profile [15], which greatly eases the implementation of the designed structures and facilitates corresponding applications. Different design methods of acoustic metagratings were reported in recent years, including drilling cavities in a rigid surface [21], nonlocal couplings between neighboring cells [22-24], meta-atoms exhibiting large Willis coupling [25,26], the integer parity of propagation number inside the metagrating [27,28], fine-tuned losses in non-Hermitian acoustic gratings [29,30], and underwater metagratings with various functionalities [31,32]. It is noted that all these metagratings work in a reciprocal way because all of them are even-symmetric under time-reversal, and the realization of nonreciprocal acoustic metagrating is still an open question, and nonreciprocal devices such as acoustic circulators are yet to be demonstrated in the context of metagrating.

In the viewpoint of applications, nonreciprocal acoustic directional couplers, which can transmit power over desired directions, are always highly desired in many realistic scenarios. Among a large variety of acoustic couplers such as power dividers, power combiners, and phase shifters, the so-called circulators, are probably the most important ones. It is really beneficial for potential applications if such nonreciprocal circulators can be realized via wavelength-thick structures that do not require high power levels or impractically large volumes, and do not rely on weak magnetoelastic effect. In the following, we describe the design, operation, and performance of innovative acoustic circulators based on intelligently designed metagratings. We show that a simply-structured metagrating, consisting of a few hard cylinders and a steady fluid flow velocity field only, can be capable of transmitting acoustic signals in a highly nonreciprocal way with unitary efficiency at the desired working frequency. Through full-wave numerical simulations, we demonstrate the circulator response at all ports, confirming its nonreciprocal operation and circulation of the input acoustic signal over a broad range of flow velocity. Since the magnitude of the biased fluid flow velocity can be electrically controlled, our design provides a feasible degree of tunability, with the possibility of switching from reciprocal to nonreciprocal operation, and the possibility of reversing the handedness of the circulator by simply changing the direction of the background flow velocity.

## II. DESIGN OF METAGRATING-BASED NONRECIPROCAL DEVICES

Let us consider a metagrating-based circulator for waterborne sound with giant linear nonreciprocal response. Each unit cell of the metagrating is composed of two hard cylinders with radii $r_1$ and $r_2$, respectively, which are located at positions $(x_1, y_1)$ and $(x_2, y_2)$ in front of a sound hard plane, with $d$ being the periodicity of the metagrating, as shown schematically in Fig. 1(a).

Let us consider a uniform inflow field $\vec{v}_0 = v_0 \hat{e}_x$ as an example, with $\hat{e}_x$ pointing to the *x*-direction. Different from previous works [3], here we impart a linear instead of a circular motion of fluid. When hard cylinders are placed within this uniform inflow $\vec{v}_0$, due to the fluid-structure interaction, the resultant background fluid field $\vec{v}_{bg}$ will be different from the inflow field $\vec{v}_0$ in regions around the cylinders, although $\vec{v}_{bg} = \vec{v}_0$ in regions far away from the cylinders. Actually, according to fluid dynamics, the normal component of $\vec{v}_{bg}$ must vanish at each cylinder's surface. To be more specific, when cylinders with position coordinates $(x_i, y_i)$ and radius $r_i$ are embedded in the inflow field $\vec{v}_0 = v_0 \hat{e}_x$, the resultant background flow velocity $\vec{v}_{bg} = (v_x, v_y)$ is given by

$$\begin{cases} v_x = v_0 \left(1 - \sum_{i=1}^{m} \frac{r_i^2}{\xi_i^2} \cos(2\beta_i)\right) \\ v_y = -v_0 \sum_{i=1}^{m} \frac{r_i^2}{\xi_i^2} \sin(2\beta_i) \end{cases}, \quad (1)$$

where $\xi_i = \sqrt{(x - x_i)^2 + (y - y_i)^2}$ and $\beta_i = arctan2(y - y_i, x - x_i)$ are, respectively, the radial and angular coordinates measured with respect to the *i*-th cylinder.

In Fig. 1(b), we plot the background flow velocity field $\vec{v}_{bg}$ for a metagrating composed of two rigid cylinders embedded in a uniform inflow $\vec{v}_0 = 50 \ (m/s) \ \hat{e}_x$, with colors and arrows representing the magnitudes and directions of $\vec{v}_{bg}$, respectively. Corresponding streamlines are plotted in Fig. 1(c), where water flows around the circular cylinders are seen. We note that $v_0 = 50 \ m/s \ll c_0$, with $c_0 = 1490 m/s$ being the sound velocity in static water.

On the other hand, according to the theory of diffraction, a plane wave incident onto the metagrating will be reflected into several plane wave beams of different diffraction orders. For a plane wave with angular frequency $\omega_0$ and incident angle $\theta_{in}$, when there is no water flow, the reflected angle $\theta_{re}$ is determined via

$$k_x = k_0 sin\theta_{re} = k_0 sin\theta_{in} + n\frac{2\pi}{d} \qquad (2)$$

where $k_0 = \omega_0/c_0$ is the wave-number in water, and $n$ is an integer representing the diffraction order. For the manipulation of propagating diffraction waves, we need to ensure that $|sin\theta_{re}| \leq 1$, which means that $n$ can take only a few possible values.

When there is no fluid motion, as shown in a recent work [31], we can design the acoustic metagrating so that the scattered waves are engineered in a highly controlled manner: for a plane wave incident from any input port, we can distribute the reflected power among all output ports in an arbitrary ratio. That is to say, the scattering parameter $S_{ij}$ ($i \neq j$) may take any value with a magnitude between 0 and 1 through a proper design of the metagrating. In this stage, the law of specular reflection is already broken, but the reciprocity still holds: we always have $S_{ij} = S_{ji}$ no matter what their values are, i.e., the scattering matrix $S$ is a symmetrical one for a reciprocal system.

But when there is a background fluid velocity field $\vec{v}_{bg}$, the situation is more complex and inherently different. The governing equation is not acoustic wave equation anymore, but becomes the so-called linearized equation for potential flow [33,34],

$$-\frac{\rho_0}{c_0^2}i\omega\left(i\omega\phi + \vec{v}_{bg} \cdot \nabla\phi\right) + \nabla \cdot \left(\rho_0\nabla\phi - \frac{\rho_0}{c_0^2}\left(i\omega\phi + \vec{v}_{bg} \cdot \nabla\phi\right)\vec{v}_{bg}\right) = 0 \qquad (3)$$

where $\vec{v}_{bg}$ is the background flow velocity field, $\phi$ represents the velocity potential of acoustic wave, and $\rho_0 = 1000 kg/m^3$ is the mass density of water. In regions away from the embedded cylinders, $\vec{v}_{bg} = \vec{v}_0$ is a homogeneous field, and the sound velocity measured in the (fixed) laboratory frame of reference (in which the fluid flows) is changed from $c_0$ to $c = c_0(1 + Mcos\alpha)$, where $M = v_0/c_0$ is the Mach number, and $\alpha$ is the angle between $\vec{v}_0$ and the propagation direction of plane wave [35]. This is the Doppler effect, so that the sound velocities parallel and antiparallel to $\vec{v}_0$ are $c_0 + v_0$ and $c_0 - v_0$, respectively. As a result, the wave-number $k$ is given by

$$k = \frac{k_0}{1+Mcos\alpha}. \qquad (4)$$

For a moving fluid, the wave-number along the x-direction is no longer $k_0 sin\theta$ due to the biased fluid motion, and we are facing a more complex situation than Eq. (2). With a nonzero background flow velocity $\vec{v}_{bg}$, a nonreciprocal response is possible, and the scattering matrix $S$ can be asymmetrical. The ultimate value of each scattering parameter $S_{ij}$ is determined by the summation of all scattered waves by every structural unit of the metagrating in a steady flow of $\vec{v}_{bg}$, plus the complex multiple-scattering effect between each component.

In this article, we will show for a proper flow velocity field and through an intelligent design, giant nonreciprocity with unitary efficiency between different ports/channels of acoustic metagratings can be achieved via scattering interference of sound. In our study, the maximum value of inflow velocity is not larger than $50\ m/s$, thus the Mach number ($M \leq 0.03$) is quite low, making it practically feasible.

To demonstrate nonreciprocal response effects, we begin with a three-port circulator, as shown schematically in Fig. 1(a). We require the following nonreciprocal response between the ports: $1 \to 2$, $2 \to 3$, and $3 \to 1$. That is to say, a plane wave incident from port 1 is totally and exclusively reflected to port 2. From port 2, however, power only transmits to port 3 and, likewise, from 3 to 1. Such acoustic device is characterized with an asymmetric scattering matrix $S$:

$$S = \begin{bmatrix} 0 & 0 & 1 \\ 1 & 0 & 0 \\ 0 & 1 & 0 \end{bmatrix}, \qquad (5)$$

which implies the inherent nonreciprocity: $S_{ij} \neq S_{ji}$. Since an isolator can be viewed as a subsystem of a circulator and can be readily obtained by impedance matching one of the circulator ports [3], we consider circulators as building blocks for various nonreciprocal acoustic devices, and in this article focus on its implementation via metagratings.

Obviously, the required unitary nonreciprocal response, i.e., Eq. (5), can be achieved only when the background flow velocity field $\vec{v}_{kg}$ and geometrical parameters (such as size and position) of the cylinders satisfy adequate conditions. To construct an appropriate metagrating for the desired nonreciprocal response is a typical inverse-design problem, and it is not trivial to find a satisfying solution.

If we adopt traditional design methods to realize the desired nonreciprocal effect,

we usually begin with the scattering matrix of a single cylinder. For cylinders with radii comparable to the wavelength and embedded in a steady flow velocity field, however, we have to expand the scattered acoustic waves in terms of very complicated cylindrical harmonic functions (including higher-order Bessel and Hankel functions), which makes the situation even worse when we have to consider the coherent addition of all scattering waves from an infinitely periodical array of cylinders, not to mention that all multiple-scattering effects between different harmonic components should be fully taken into consideration [36,37]. Thus, such direct design method demands extensive experience of analytic modeling such as the multiple-scattering theory, being often time-consuming and low efficient, and even practically impossible if very complex nonreciprocal response is required.

Therefore, we adopt a different approach, i.e., an intelligent optimization method [19,20,23,31,38] based on the genetic algorithm (GA) for this inverse problem, and set the structural parameters of the metagrating as optimization variables, and the reflection efficiency (such as $S_{21}$, $S_{32}$, and $S_{13}$) between different ports as the optimization target. It turns out that by utilizing the GA-assisted optimization process, we can quickly find a set of design parameters that can satisfyingly produce the desired nonreciprocal functionality. We verify the results with full-wave numerical simulations by using COMSOL Multiphysics, a finite-element-based soft package. Nearly perfect nonreciprocal functionalities are unambiguously demonstrated over a broad range of inflow velocity profiles for simply-structured metagratings, for both three-port and four-port circulators, with either sound-soft or sound-hard boundaries.

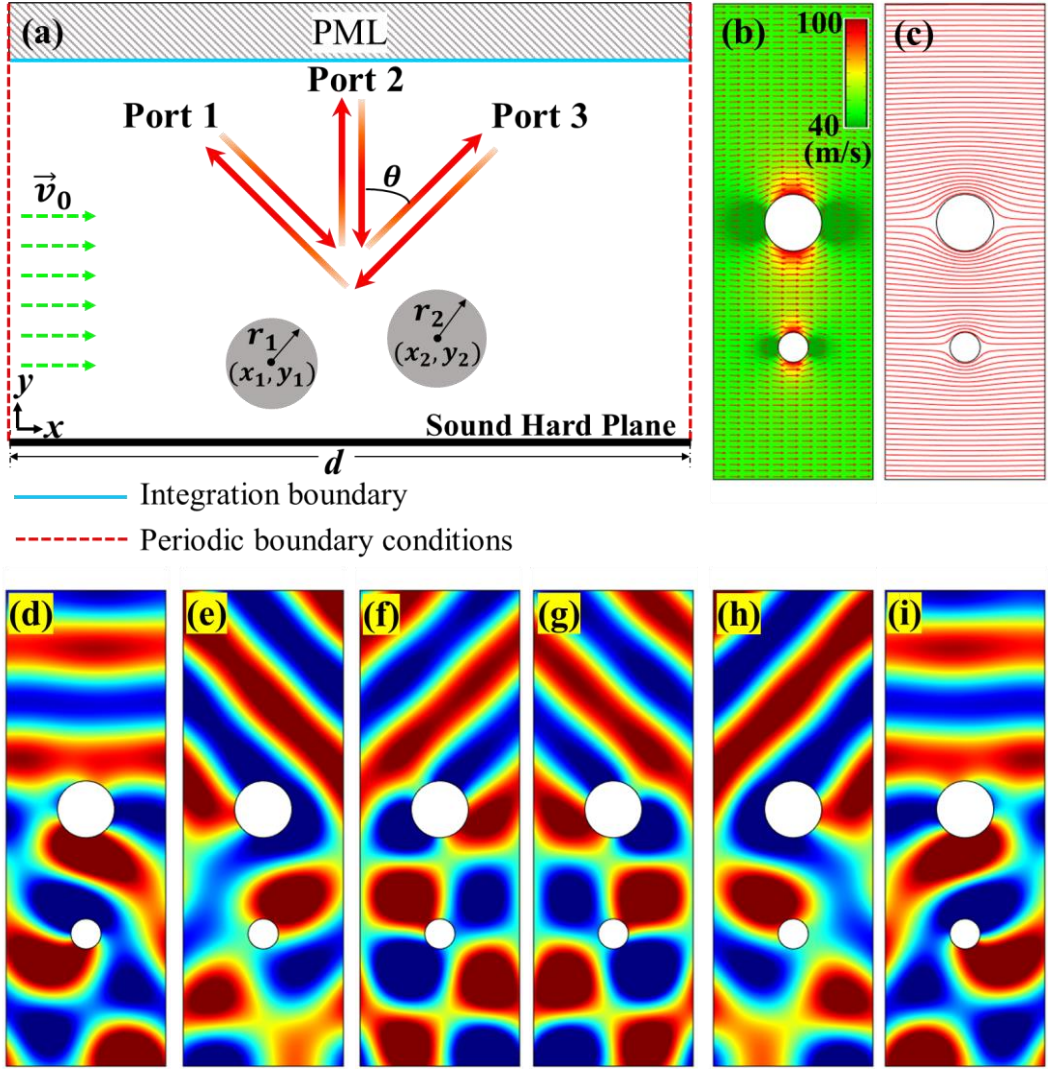

**Figure 1.** An acoustic metagrating with nonreciprocal response between three ports. (a) Unit cell configuration of the metagrating, where in front of a sound hard plane there are two rigid cylinders embedded in a steady water inflow $\vec{v}_0 = 50\ (m/s)\ \hat{e}_x$. (b) Colors and arrows represent the magnitude and direction of the corresponding background flow velocity $\vec{v}_{bg}$. (c) Streamlines for $\vec{v}_{bg}$. (d), (e) and (f) show the *scattered* pressure fields by a metagrating with $x_1 = x_2$ when the plane waves are incident from ports 1, 2, and 3, respectively, with corresponding diffraction angle $\theta = 45°$. Nonreciprocal reflection behavior between the ports is obtained: $1 \rightarrow 2$, $2 \rightarrow 3$, and $3 \rightarrow 1$. (g), (h), and (i) show the corresponding scattered fields after the direction of $\vec{v}_0$ is *reversed* from a positive *x*-direction to a negative one, and the nonreciprocal behavior becomes $1 \rightarrow 3$, $3 \rightarrow 2$, and $2 \rightarrow 1$.

## II. RESULTS

### A. Three-port acoustic circulators based on metagratings

To demonstrate the possibility of realizing giant nonreciprocity through proposed metagrating structures, we assume the working frequency is 2,000 Hz, corresponding a wavelength of 74.5 cm in water. For the three-port circulator shown in Fig. 1(a), we specify the diffraction angle $\theta = 45°$ as an example, and we note that $\theta$ can take a different value that depends on realistic demands. The periodicity $d$ of the metagrating is given by $d = \lambda/sin\theta = 105.36\ cm$. Since the three ports are symmetrical with respect to the $y$-axis, and the inflow velocity $\vec{v}_0$ is along the positive $x$-direction, we speculate that in order to fulfill the required nonreciprocal response, the positions of two cylinders could have some spatial symmetry. As an example, we put both cylinders at the middle line of the unit cell ($x_1 = x_2 = d/2$), as shown in Fig. 1(c)-(h). After applying the GA-assisted optimization method, we find the following optimized parameters for the two cylinders: $r_1$ = 9.97 cm, $r_2$ = 18.84 cm, $y_1$ = 87.89cm, and $y_2$ = 170.65 cm. Then we perform full-wave numerical simulations for verification. We plot the scattered pressure waves in Figs. 1(d), (e) and (f), respectively, when plane waves are incident from ports 1, 2, and 3. Obviously, the scattered waves are plane waves propagating towards ports 2, 3, and 1, respectively. The desired giant nonreciprocity is indeed achieved.

After we change the direction of inflow velocity $\vec{v}_0$ from a positive $x$-direction to a negative one while keeping all other conditions unchanged, we find that the handedness of the circulator is reversed. The nonreciprocal reflection response becomes: $1 \to 3$, $3 \to 2$, and $2 \to 1$, and the scattering matrix turns into

$$S = \begin{bmatrix} 0 & 1 & 0 \\ 0 & 0 & 1 \\ 1 & 0 & 0 \end{bmatrix}. \qquad (6)$$

The corresponding scattered pressure fields are plots in Figs. 1(g), (h), and (i), respectively, and nonreciprocal response with unitary efficiency can be recognized through the perfect planar wave-front patterns exhibited by the scattered waves.

As a comparison, we also plot the scattered pressure fields in Figs. 2(a)-(c) when there is no background water flow. Strong interference patterns are observed, which means that the reflected wave consists of more than one diffraction order, and the power flow is split between all three ports. Thus, the unitary reflection efficiency demonstrated in Figs. 1(d)-(i) is the combined result of a proper inflow velocity $\vec{v}_0$ and the complex multiple-scattering effect between the all cylinders with adequate parameters. Since the magnitude of the fluid velocity can be externally controlled (e.g., via electric signals, please see Section II.C for details), our design provides a feasible degree of tunability, exhibiting the possibility of switching from reciprocal to nonreciprocal operation, as well as the tunability of reversing the handedness of the circulator by simply changing the direction of the background flow velocity.

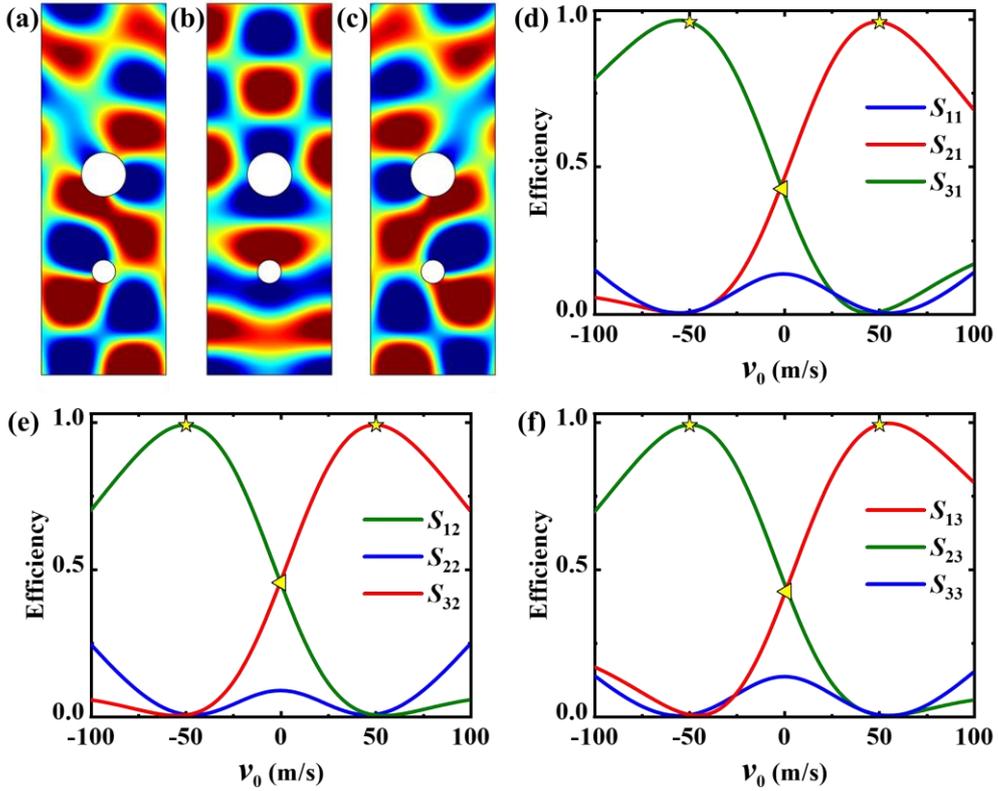

**Figure 2.** The influence of water inflow $\vec{v}_0$ on the nonreciprocal response. (a), (b), and (c) show the scattered pressure fields by the metagrating when there is no water flow. (d), (e), and (f) show how the diffraction efficiencies vary as the inflow velocity $v_0$ changes.

To gain further insights into the response of the proposed metagrating, we study the influence of the magnitude of inflow velocity $\vec{v}_0$ on the diffraction efficiency between all three ports. The results are shown in Figs. 2(d), (e) and (f), respectively, when planes are incident from ports 1, 2, and 3. Let us take Fig. 2(d) as an example, where a plane wave is incident from port 1. When there is no water flow ($v_0 = 0\ m/s$), the power flow is split between three output ports, as shown by the yellow triangle in Fig. 2(d), and the response of the metagrating is fully reciprocal. As the inflow velocity $v_0$ increases from zero along the positive-$x$ direction, $S_{31}$ (green curve in Fig. 2(d)) gradually goes down to zero, whereas $S_{21}$ (red curve) monotonically increases to unity at the optimized velocity magnitude ($v_0 = +50\ m/s$), as shown by the yellow star. At this point, all scattered waves interfere constructively with each other so that a perfect reflection to the desired port (i.e., port 2) is achieved. Simultaneously, destructive interference is obtained at other ports, and no power is directed to the unwanted ports.

Beyond this optimal value, $S_{31}$ increases again, whereas $S_{21}$ decreases, which is similar to the phenomenon observed in [3]. The proposed nonreciprocal effect is robust to fluctuations in the background velocity, and our simulations predict a large degree of isolation ratio $|S_{21}/S_{31}|$ over a broad range of flow velocities: we have $|S_{21}/S_{31}| > 10$ as long as $25\ m/s < v_0 < 76\ m/s$. Thus, the nonreciprocal reflection of acoustic signal from port 1 to port 2 is not so sensitive to the value of flow velocity, which is an advantage for various applications. A similar nonreciprocal response can be observed when $v_0$ is increased from zero along the negative-$x$ direction, or when the plane waves are incident from ports 2 and 3, as shown in Fig. 2(e) and (f).

In Fig. 3(a), (b), and (c), we show the diffraction efficiency spectra between all ports when the magnitude of $\vec{v}_0$ is varied. At the working frequency of 2,000 Hz, we obtain perfect diffraction efficiency along desired channels, i.e., $S_{21} = 99.0\%$, $S_{32} = 99.1\%$, and $S_{13} = 99.1\%$, as marked by yellow stars. As a comparison, in Fig. 3(d), (e) and (f) we show the corresponding diffraction efficiency spectra when there is no background flow, where we see no unitary reflection to the desired ports, and a fully reciprocal behavior can be verified by the identical profiles of scattering parameters: $S_{ij} = S_{ji}\ (i \neq j)$.

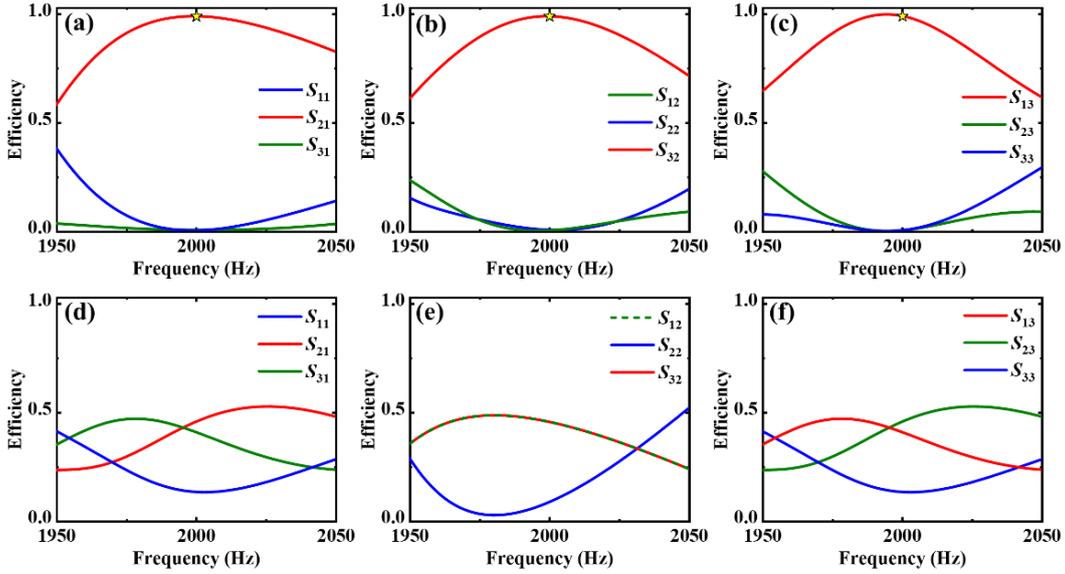

**Figure 3.** Diffraction efficiency of the three-port metagrating in Fig. 1. (a), (b), and (c) show the diffraction efficiency spectra for plane waves incident from ports 1, 2, and 3, respectively. At the operating frequency of 2,000 Hz, nearly unitary diffraction efficiencies are achieved, i.e., $S_{21}$ = 99.0%, $S_{32}$ = 99.1%, and $S_{13}$ = 99.1%, as shown by yellow stars. In comparison, (d), (e), and (f) show the corresponding diffraction efficiency spectra when there is no water flow, where a fully reciprocal behavior, i.e., $S_{ij} = S_{ji}$ ($i \neq j$), can be observed.

From Fig. 3(a)-(c), we observe that the spectra of $S_{21}$, $S_{32}$, and $S_{13}$ (red curves) are similar in profile, and all of them exhibit a relative broad peak in terms of the frequency range (64-Hz bandwidth with $S_{21}$, $S_{32}$, and $S_{13} > 0.80$). At the same time, all other diffraction efficiency spectra (green and blue curves) are highly suppressed over a relative broad frequency range around the working frequency at 2,000 Hz. Such robust behavior of diffraction efficiency $S_{ij}$ with respect to the working frequency is beneficial to nonreciprocal response for realistic applications.

We note that the nonreciprocal response of the metagrating is not limited to the rigid-cylinder configuration, but can be also realized with elastic solids such as iron, aluminum, and rubber cylinders. In addition, the two cylinders do not necessarily have any spatial symmetry, and as an example we present a metagrating with asymmetric configuration in Appendix A.

**B. Four-port acoustic circulators**

In the previous section, we demonstrate three-port acoustic circulators with unitary reflection efficiency. The same design paradigm can be generalized to more ports. In the following, we will design four-port circulators based on metagratings, where diffraction channels up to $n = \pm 2$ need to be considered, and the four ports correspond to diffraction orders with propagation angles $+\theta_1$, $+\theta_2$, $-\theta_1$, and $-\theta_2$, respectively. To make sure that higher diffraction orders ($|n| \geq 3$) are evanescent waves, the first-order reflection angle $\theta_1$ should satisfy $19.47° < \theta_1 < 30°$. We take $\theta_1 = 25°$ as an example, then $\theta_2$ can be determined as $\theta_2 = 57.67°$ according to the diffraction theory. We desire to realize a nonreciprocal response with unitary efficiency between the four ports: $1 \to 2$, $2 \to 3$, and $3 \to 4$, and $4 \to 1$, corresponding an asymmetric scattering matrix $S$:

$$S = \begin{bmatrix} 0 & 0 & 0 & 1 \\ 1 & 0 & 0 & 0 \\ 0 & 1 & 0 & 0 \\ 0 & 0 & 1 & 0 \end{bmatrix} \qquad (7)$$

Different from previous three-port circulators, we introduce a non-uniform inflow $\vec{v}_0$ for the four-port circulators, with $\vec{v}_0$ given by

$$\vec{v}_0 = \begin{cases} 50(m/s) \cdot \frac{1}{0.6d} y \hat{e}_x; & 0 \leq y \leq 0.6d \\ 50(m/s) \cdot \frac{1}{0.6d}(1.2d - y)\hat{e}_x; & 0.6d < y \leq 1.2d \end{cases}. \qquad (8)$$

At this time, a nonzero inflow field exist within the region $0 < y < 1.2d$. The triangular profile of $\vec{v}_0$ is schematically shown with green arrows in Fig. 4(a), and its magnitude $v_0$ is a linear function of the $y$-coordinate. Obviously, $v_0$ reaches its maximum $50 \, m/s$ at the plane $y = 0.6d$, while decreases to zero at $y = 0$ and $y = 1.2d$. Here we intentionally introduce a non-uniform inflow profile of $\vec{v}_0$ to demonstrate that the nonreciprocal response is not limited to uniform inflows only.

Figure 4(a) show one configuration of the megatrating with three equal-height and equal-sized hard cylinders in each unit cell. There is a sound hard plane at the bottom, with geometrical parameters explicitly marked, where $d$ = 176.28 cm is the period of metagrating. The geometric parameters of the cylinders are the optimization variables, and $S_{21}$, $S_{32}$, $S_{43}$, and $S_{14}$ are the optimization targets. Without loss of generality, we put three identical cylinders symmetrically in the unit cell, i.e., one cylinder right at the middle line, i.e., $x_2 = d/2$, and other two cylinders equidistant from the central one, i.e., $x_1 = d/2 - \Delta L$, and $x_3 = d/2 + \Delta L$. By applying the GA-based optimization method, we find the following parameters: $r = 18.11\ cm$, $y = 152.35\ cm$, and $\Delta L = 58.78\ cm$. Figures 4(c), (d), (e), and (f) show the scattered pressure fields when the plane waves are incident from ports 1, 2, 3, and 4, respectively, and nearly perfect nonreciprocal response can be recognized from the planar wave-front of the scattered wave.

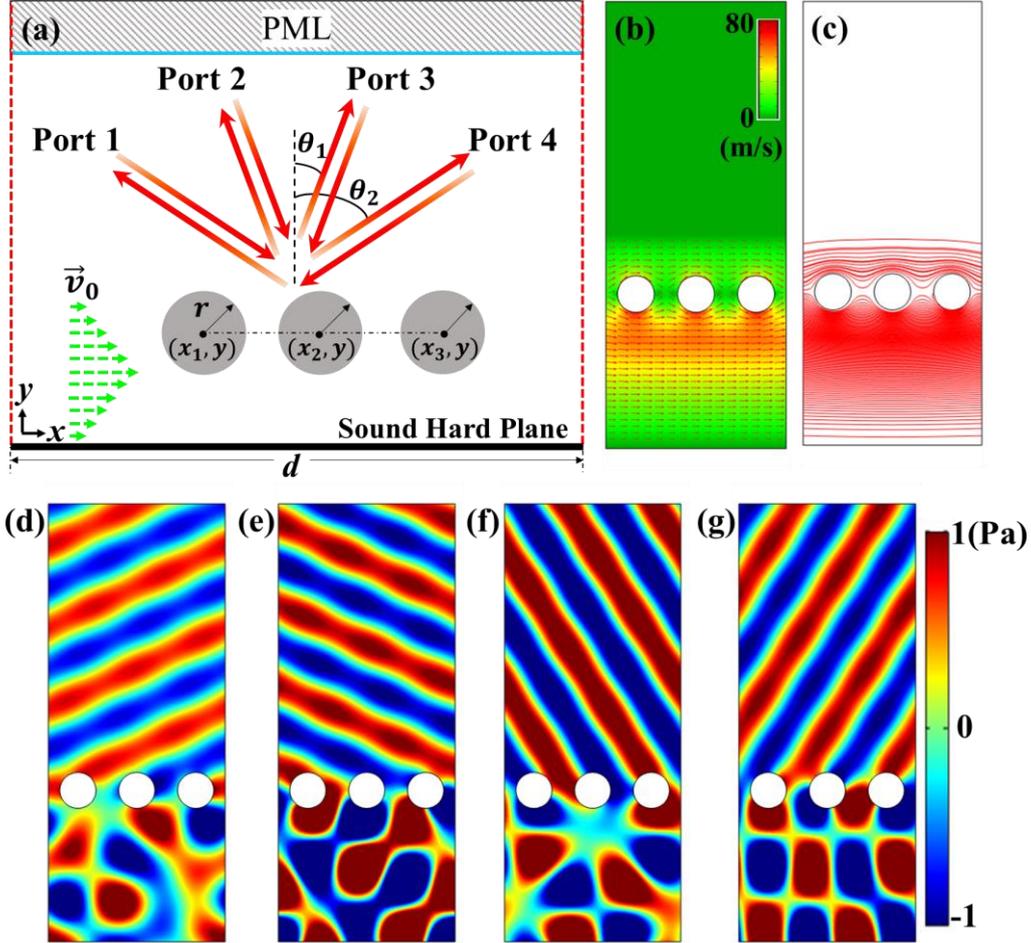

**Figure 4.** An acoustic circulator with four ports. (a) Unit cell configuration of the metagrating, where in front of a sound hard plane there are three equal-height and equal-sized hard cylinders embedded in an inflow velocity $\vec{v}_0$ as specified by Eq. (8). (b) Colors and arrows represent the magnitude and direction of the corresponding background flow velocity $\vec{v}_{bg}$. (c) Streamlines for $\vec{v}_{bg}$. (d), (e), (f), and (g) show the scattered pressure fields in water when the plane waves are incident from ports 1, 2, 3, and 4, respectively, with diffraction angles $\theta_1 = 25°$ and $\theta_2 = 57.67°$. Nonreciprocal reflection behaviors between the ports, i.e., 1→2, 2→3, 3→4, and 4→1, are seen.

Actually, in addition to the triangular profile as specified in Eq. (8), we have also studied other non-uniform inflow profiles of $\vec{v}_0$ such as a parabolic or trapezoidal profile. We find that unitary reflection efficiency can be also achieved, which shows that the nonreciprocal response is indeed robust over different inflow profiles.

We find that the handness of the four-port circulator can be also reversed if we change the direction of the inflow velocity $\vec{v}_0$. When the water flows along the negative-$x$ direction, the following nonreciprocal behaviors are obtained: $1 \rightarrow 4$, $4 \rightarrow 3$, $3 \rightarrow 2$, and $2 \rightarrow 1$, and the scattering matrix becomes

$$S = \begin{bmatrix} 0 & 1 & 0 & 0 \\ 0 & 0 & 1 & 0 \\ 0 & 0 & 0 & 1 \\ 1 & 0 & 0 & 0 \end{bmatrix}. \tag{9}$$

Thus, the four-port circulator is also a tunable nonreciprocal device that can be electrically controlled in a similar way to three-port circulators.

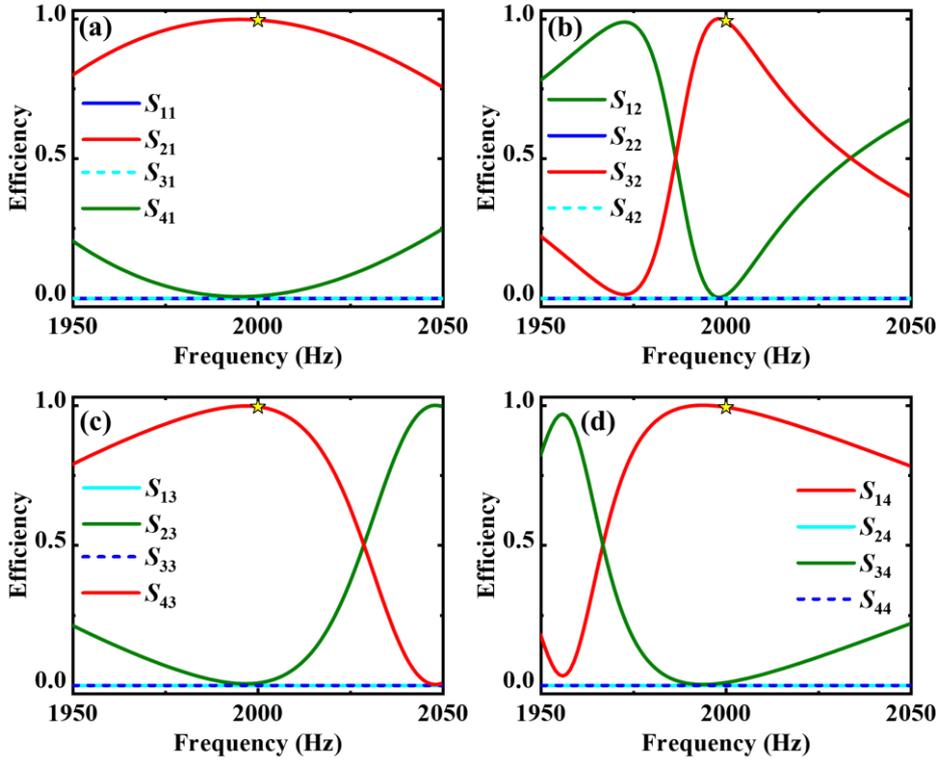

**Figure 5**. Diffraction efficiency spectra of the four-port circulator when plane waves are incident from ports 1, 2, 3, and 4, respectively, are shown in (a), (b), (c), and (d). Almost unitary diffraction efficiency, i.e., $S_{21}$ = 99.5%, $S_{32}$ = 99.3%, $S_{43}$ = 99.5%, and $S_{14}$ = 99.3% are achieved at the working frequency 2,000 Hz, respectively.

The diffraction efficiency spectra are shown in Fig. 5, corresponding to plane waves incident from ports 1, 2, 3, and 4, respectively. Nearly unitary diffraction efficiencies are achieved at the working frequency 2,000 Hz, with $S_{21} = 99.5\%$, $S_{32} = 99.3\%$, $S_{43} = 99.5\%$, and $S_{14} = 99.3\%$, as marked by yellow stars. We also observe that all blue and cyan curves are greatly suppressed and very close to zero around 2,000 Hz, with blue curves representing retro-reflection efficiencies (i.e., the reflection angle is the same as the incident angle).

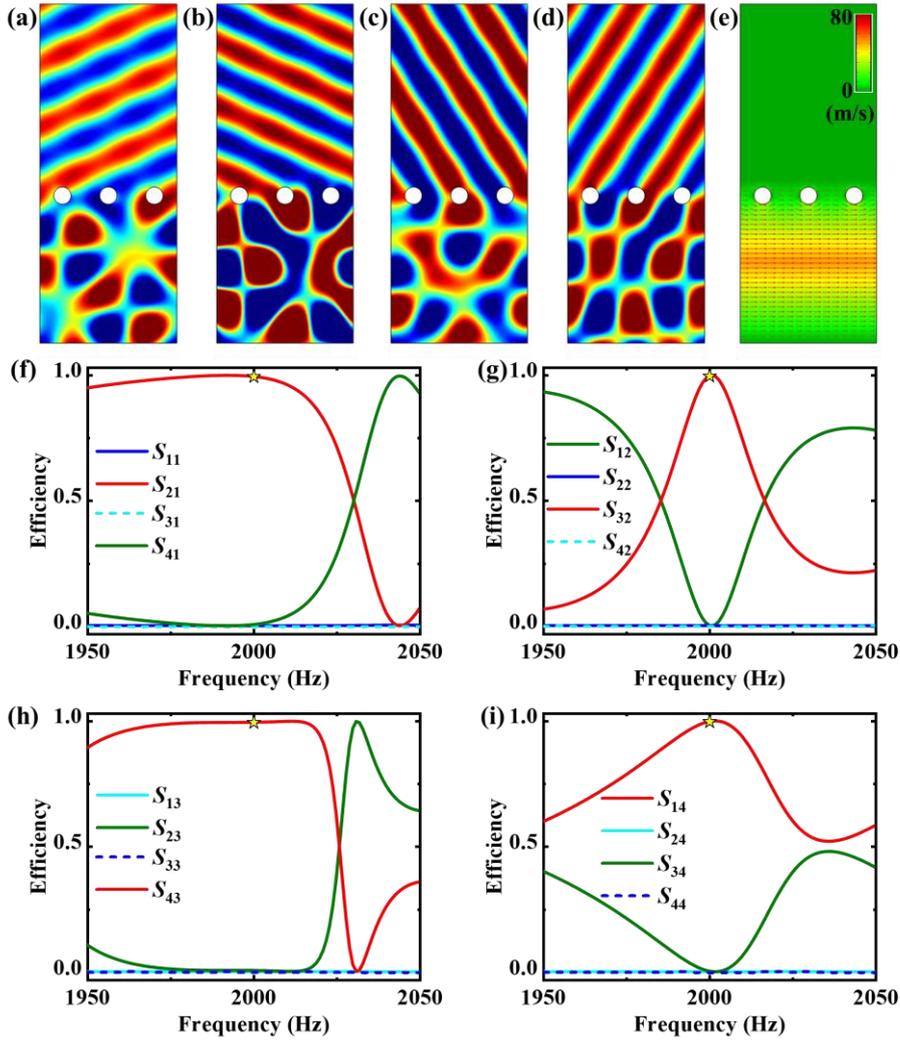

**Figure 6**. Another four-port acoustic circulator, where a sound soft plane is utilized at the bottom. (a), (b), (c), and (d) show the scattered pressure fields in water when the plane waves are incident from ports 1, 2, 3, and 4, respectively, with the reflection angle $\theta_1 = 25°$, and $\theta_2 = 57.67°$. (e), (f), (g), and (h) show the diffraction efficiency spectra for all channels when plane waves are incident from ports 1, 2, 3, and 4, respectively. Nearly unitary diffraction efficiency, i.e., $S_{21} = 99.4\%$, $S_{32} = $

99.6%, $S_{43}$ = 99.4%, $S_{14}$ = 99.7% are achieved at the working frequency 2,000 Hz.

In previous examples, a sound hard plane is used in the metagrating. In Fig. 6 we will show that a sound soft plane can also be utilized for the construction of nonreciprocal metagrating. We keep the background flow velocity field the same as that in Fig. 4, and put three identical cylinders symmetrically in the unit cell. After applying the GA-assisted method, we find the following optimized geometric parameters: $r = 11.18\ cm$, $y = 191.71\ cm$, $x_2 = d/2$, and $x_{3,1} = d/2 \pm 59.2$ cm. Figures 6(a), (b), (c), and (d) show the scattered pressure fields in water when plane waves are incident from ports 1, 2, 3, and 4, respectively. Corresponding diffraction efficiency spectra are shown in Figs. 6(e)-(h). Almost perfect diffraction efficiencies $S_{21} = 99.4\%$, $S_{32} = 99.6\%$, $S_{43} = 99.4\%$, $S_{14} = 99.7\%$ are achieved at the working frequency 2,000 Hz, as marked by yellow stars.

## C. METHODS FOR NUMERICAL SIMULATIONS AND EXPERIMENTAL REALIZATIONS.

In the numerical simulations, we use the Linearized Navier-Stokes (frequency domain) module in COMSOL Multiphysics to study the nonreciprocal response. In this work we consider non-viscous fluids, and set the fluid viscosity (dynamic viscosity and bulk viscosity), thermal conductivity, and the viscous dissipation function to zero. Therefore, the resultant dynamic equation is reduced to a linearized one for the potential flow, as expressed by Eq. (3).

According to fluid dynamics, the background fluid flow velocity field $\vec{v}_{bg}$ is determined by the inflow velocity profile $\vec{v}_0$ as well as the positions and sizes of the embedded cylinders. After a proper setup of $\vec{v}_{bg}$, we can excite the metagrating by an incident acoustic plane wave. The reflection efficiencies for various ports/channels can be calculated by integrating corresponding pressure fields at the upper boundary of the simulation domain, as marked by the blue curves in Fig.1(a) & Fig.4(a). Due to the Doppler effect, the reflection angles for various ports are dependent on whether the background velocity field $\vec{v}_{bg}$ is zero on the integration boundary (please refer to Appendix B for details). The Floquet periodic boundary conditions are applied to the left and right boundaries (marked by red dashed lines) of the simulation domain, and the upper boundary is adjacent to a perfect matching layer (PML) to absorb outgoing acoustic waves.

Finally, we would like to discuss two possible methods that could be used for the control of fluid velocity. In the first method, the velocity of the inflow fluid can be controlled by using an electric piston pump, as proposed in Ref. [39]. The water is firstly pumped into a sink through a rubber tube, where the flow velocity in the tube is electrically controlled by a piston pump connected with the tube. After going through the sink, the water flows into a rectangular channel uniformly, where cylinders are placed and the resultant background fluid velocity field is generated. A piece of cover is placed on top of the samples to enclose the channel. Thus, the flow velocity in the channel is determined by the flow velocity in the rubber tube, and eventually controlled by the pump.

Since the background water velocity is nothing but the relative speed between the cylinders and water background, in the second method we can keep water still, but let all equipment's and cylinders (mounted on a translating platform) move simultaneously at the same speed. In this way, we can obtain a uniform background flow velocity by the method of relative motion. Actually, a similar method was used recently in Ref. [40].

## IV. CONCLUSION

We propose to realize nonreciprocal circulators from simply-structured acoustic metagratings, which consist only of a few solid cylinders and a steady fluid flow velocity field. These acoustic circulators can transport wave signals in a highly nonreciprocal way with unitary efficiency at the desired working frequency, and the circulation of the signal can be operated over a broad range of velocity magnitude and with different velocity field profiles. Our design provides a feasible degree of tunability, exhibiting the possibility of switching from reciprocal to nonreciprocal operation, and even reversing the handedness of the circulator through changing the magnitude or direction of the background flow velocity.

The proposed nonreciprocal effect is largely tunable and scalable from audible to ultrasonic frequencies and provides an alternative solution to realize acoustic switching, rectification, isolation, and circulation with metagratings. We anticipate that this concept may open new directions in the research of nonreciprocal acoustic devices, including underwater communication, energy harvesting, and acoustic sensing.

## ACKNOWLEDGMENS

This work is supported by National Natural Science Foundation of China (11574087) and Guangdong Basic and Applied Basic Research Foundation (2021A1515010322).

## REFERENCES

1. H. Nassar, B. Yousefzadeh, R. Fleury, M. Ruzzene, A. Alù, C. Daraio, A. N. Norris, G. Huang, and M. R. Haberman, Nonreciprocity in acoustic and elastic materials, Nat. Rev. Mater. **5**, 667 (2020).
2. A. Alù, Magnet-Free Nonreciprocity, Proceedings of the IEEE **108**, 1682 (2020).
3. R. Fleury, D. L. Sounas, C. F. Sieck, M. R. Haberman, and A. Alù, Sound Isolation and Giant Linear Nonreciprocity in a Compact Acoustic Circulator, Science **343**, 516 (2014).
4. C. P. Wiederhold, D. L. Sounas, and A. Alù, Nonreciprocal acoustic propagation and leaky-wave radiation in a waveguide with flow, J. Acoust. Soc. Am. **146**, 802 (2019).
5. F. Zangeneh-Nejad and R. Fleury, Acoustic rat-race coupler and its applications in non-reciprocal systems, J. Acoust. Soc. Am. **146**, 843 (2019).
6. R. Sasaki, Y. Nii, Y. Iguchi, and Y. Onose, Nonreciprocal propagation of surface acoustic wave in Ni/LiNbO$_3$, Phys. Rev. B **95**, 020407 (2017).
7. R. Verba, I. Lisenkov, I. Krivorotov, V. Tiberkevich, and A. Slavin, Nonreciprocal surface acoustic waves in multilayers with magnetoelastic and interfacial Dzyaloshinskii-Moriya interactions, Phys. Rev. Appl. **9**, 064014 (2018).
8. D.-D. Dai and X.-F. Zhu, An effective gauge potential for nonreciprocal acoustics, Europhys. Lett. **102**, 14001 (2013).
9. R. Fleury, D. L. Sounas, and A. Alù, Subwavelength ultrasonic circulator based on spatiotemporal modulation, Phys. Rev. B **91**, 174306 (2015).
10. C. Shen, X. Zhu, J. Li, and S. A. Cummer, Nonreciprocal acoustic transmission in space-time modulated coupled resonators, Phys. Rev. B **100**, 054302 (2019).
11. B. Liang, X. S. Guo, J. Tu, D. Zhang, and J. C. Cheng, An acoustic rectifier, Nat. Mater. **9**, 989 (2010).

12. N. Boechler, G. Theocharis, and C. Daraio, Bifurcation-based acoustic switching and rectification, Nat. Mater. **10**, 665 (2011).
13. B.-I. Popa and S. A. Cummer, Non-reciprocal and highly nonlinear active acoustic metamaterials, Nat. Commun. **5**, 3398 (2014).
14. L. Shao, W. Mao, S. Maity, N. Sinclair, Y. Hu, L. Yang, and M. Lončar, Non-reciprocal transmission of microwave acoustic waves in nonlinear parity–time symmetric resonators, Nat. Electron. **3**, 267 (2020).
15. Y. Ra Di, D.L. Sounas, and A. Alù, Metagratings: Beyond the limits of graded metasurfaces for wave front control, Phys. Rev. Lett. **119**, 67404 (2017).
16. A. Epstein and O. Rabinovich, Unveiling the properties of metagratings via a detailed analytical model for synthesis and analysis, Phys. Rev. Appl. **8**, 54037 (2017).
17. H. Chalabi, Y. Ra'Di, D.L. Sounas, and A. Alù, Efficient anomalous reflection through near-field interactions in metasurfaces, Phys. Rev. B **96**, 75432 (2017).
18. Z. Tagay and C. Valagiannopoulos, Highly selective transmission and absorption from metasurfaces of periodically corrugated cylindrical particles, Phys. Rev. B **98**, 115306 (2018).
19. A. Díaz-Rubio and S.A. Tretyakov, Acoustic metasurfaces for scattering-free anomalous reflection and refraction, Phys. Rev. B **96**, 125409 (2017).
20. J. Li, C. Shen, A. Díaz-Rubio, S.A. Tretyakov, and S.A. Cummer, Systematic design and experimental demonstration of bianisotropic metasurfaces for scattering-free manipulation of acoustic wavefronts, Nat. Commun. **9**, 1 (2018).
21. D. Torrent, Acoustic anomalous reflectors based on diffraction grating engineering, Phys. Rev. B **98**, 060101 (2018).
22. H. Ni, X. Fang, Z. Hou, Y. Li, and B. Assouar, High-efficiency anomalous splitter by acoustic meta-grating, Phys. Rev. B **100**, 104104 (2019).
23. Z. Hou, X. Fang, Y. Li, and B. Assouar, Highly efficient acoustic metagrating with strongly coupled surface grooves, Phys. Rev. Appl. **12**, 034021 (2019).
24. L. Quan and A. Alù, Passive Acoustic Metasurface with Unitary Reflection Based on Nonlocality, Phys. Rev. Appl. **11**, 054077 (2019).
25. L. Quan, Y. Ra'di, D. L. Sounas, and A. Alù, Maximum Willis Coupling in Acoustic Scatterers, Phys. Rev. Lett. **120**, 254301 (2018).
26. Y. K. Chiang, S. Oberst, A. Melnikov, L. Quan, S. Marburg, A. Alù, and D.A. Powell, Reconfigurable Acoustic Metagrating for High-Efficiency Anomalous Reflection, Phys. Rev. Appl. **13**, 064067 (2020).
27. Y. Fu, C. Shen, Y. Cao, L. Gao, H. Chen, C.T. Chan, S.A. Cummer, and Y. Xu, Reversal of transmission and reflection based on acoustic metagratings with integer parity design, Nat. Commun. **10**, 1 (2019).
28. Y. Fu, Y. Cao, and Y. Xu, Multifunctional reflection in acoustic metagratings with simplified design, Appl. Phys. Lett. **114**, 53502 (2019).
29. Y. Yang, H. Jia, Y. Bi, H. Zhao, and J. Yang, Experimental Demonstration of an Acoustic Asymmetric Diffraction Grating Based on Passive Parity-Time-Symmetric Medium, Phys. Rev. App.**12**, 034040 (2019).
30. Y. Yang, H. Jia, S. Wang, P. Zhang, and J. Yang, Diffraction control in a non-Hermitian acoustic grating, Appl. Phys. Lett. **116**, 213501 (2020).
31. L. Fan and J. Mei, Metagratings for Waterborne Sound: Various Functionalities Enabled by an

Efficient Inverse-Design Approach, Phys. Rev. Appl. **14**, 044003 (2020).
32. J. He, X. Jiang, D. T, and W. Wang, Appl. Experimental demonstration of underwater ultrasound cloaking based on metagrating, Appl. Phys. Lett. **117**, 091901 (2020).
33. O. A. Godin, Reciprocity and energy theorems for waves in a compressible inhomogeneous moving fluid, Wave Motion **25**, 143 (1997).
34. L. M. B. C. Campos, On 36 Forms of the Acoustic Wave Equation in Potential Flows and Inhomogeneous Media, Appl. Mech. Rev. **60**, 149 (2007).
35. L. D. Landau and E. M. Lifshitz, Fluid Mechanics (2nd Edition), Pergamon Press, 1987.
36. J. Mei, Z. Liu, J. Shi, and D. Tian, Theory for elastic wave scattering by a two-dimensional periodical array of cylinders: An ideal approach for band-structure calculations, Phys. Rev. B **67**, 245107 (2003).
37. C. Qiu, Z. Liu, J. Mei, and M. Ke, The layer multiple-scattering method for calculating transmission coefficients of 2D phononic crystals, Solid State Commun. **134**, 765 (2005).
38. A. Håkansson and J. Sánchez-Dehesa, Acoustic lens design by genetic algorithms, Physical Review B **70**, 214302 (2004).
39. F. Tay, Y. Zhang, H. Xu, H. Goh, Y. Luo, and B. Zhang, A metamaterial-free fluid-flow cloak, arXiv preprint arXiv:1908.07169 (2019).
40. N.F. Declercq, L. Chehami, and R.P. Moiseyenko, The transmission spectrum of sound through a phononic crystal subjected to liquid flow, Appl. Phys. Lett. **112**, 24102 (2018).